# Comprehensive evidence of lasing from a 2D material enabled by a dual-resonance metasurface

**Isabel Barth[1,+], Manuel Deckart[1,+], Donato Conteduca[1], Guilherme S Arruda[2], Zeki Hayran[3], Sergej Pasko[4], Simonas Krotkus[4], Michael Heuken[4], Francesco Monticone[3], Thomas F Krauss[1], Emiliano R Martins[2], and Yue Wang[1,*]**

[1] School of Physics, Engineering and Technology, University of York, York, YO10 5DD, United Kingdom
[2] School of Engineering, Department of Electrical and Computer Engineering, University of Sao Paulo, Sao Carlos-SP 13566-590, Brazil
[3] School of Electrical and Computer Engineering, Cornell University, Ithaca, New York,14853, USA
[4] AIXTRON SE, DornkaulstraBe. 2, 52134 Herzogenrath, Germany
* Corresponding author email: yue.wang@york.ac.uk
+ these authors contributed equally to this work

## ABSTRACT

Semiconducting transition metal dichalcogenides (TMDs) have gained significant attention as a gain medium for nanolasers, owing to their unique ability to be easily placed and stacked on virtually any substrate. However, the atomically thin nature of the active material in existing TMD nanolasers presents a challenge, as their limited output power makes it difficult to distinguish between true laser operation and other "laser-like" phenomena. Here, we present comprehensive evidence of lasing from a CVD-grown tungsten disulphide ($WS_2$) monolayer. The monolayer is placed on a dual-resonance dielectric metasurface with a rectangular lattice designed to enhance both absorption and emission, resulting in an ultralow threshold operation (threshold <1 W/cm$^2$). We provide a thorough study of the laser performance at room temperature, paying special attention to directionality, output power, and spatial coherence. Notably, our lasers demonstrated a coherence length of over 30 μm, which is several times greater than what has been reported for 2D material lasers so far. Our realisation of a single-mode laser from a wafer-scale CVD-grown monolayer presents exciting opportunities for integration and the development of novel applications.

## Introduction

Since the first demonstration of room-temperature vertical-cavity surface-emitting lasers (VCSELs) in 1988[1], tremendous progress has been made in the field of semiconductor nanolasers, driven by the desire to reduce the lasing threshold and open up novel functionalities. Major breakthroughs are frequently linked to the discovery of new materials that can amplify light and advancements in the fabrication processes of nanodevices. Recently, new opportunities have arisen with the emergence of two-dimensional (2D) layered van der Waals (vdW) materials, especially the transition-metal dichalcogenide (TMD) semiconductors[2, 3] and their heterostructures[4,5].

Light sources based on 2D TMDs are particularly attractive due to their ability to be placed on a wide range of substrates, including wearables. The light emission properties of these materials are determined by excitons and trions, which result from strong Coulomb interactions[6]. In particular, the excitons exhibit a strong binding energy of typically hundreds of meV, one or two orders of magnitude higher than, for example, GaAs quantum wells, which leads to sub-nanosecond spontaneous emission lifetimes and stable room-temperature operation. By stacking different monolayers of TMD vertically, the formation of vdW-bonded heterostructures is possible without encountering the typical "lattice mismatch" problem. This unique capability reveals extraordinary phenomena and enables the development of innovative optoelectronic devices, such as interlayer excitonic lasers that emit light at longer wavelengths into the infrared range[7, 8].

In order to realise these exciting opportunities, a number of issues need to be resolved: (a) limited absorption at the pump wavelength due to the nanometer-thin gain medium; (b) small mode overlap with the gain medium at the lasing wavelength for the same reason; (c) the gain material's lateral size is often limited, with mechanical exfoliation being the most common method for obtaining high-quality TMD materials until recently, which typically results in an active area limited to the micron-scale. Here, we address these issues by employing a design to resonantly enhance both the absorption at the pump wavelength and the emission at the lasing wavelength, and by utilising high-quality, large-area TMD monolayers grown by metal-organic chemical vapor deposition (MOCVD).

Resonant enhancement has been used before, e.g. using high-Q whispering-gallery modes[9, 10], distributed- Bragg-reflector cavities[11] , or photonic crystal cavities[12], but mostly at the emission wavelength only. Optical bound states in the continuum (BICs)[13] have also been adopted for low-threshold lasing[14–16]. Along similar lines, a 'pseudo-BIC' $WS_2$ laser with a compact active region with external reflectors was recently demonstrated[17]. It should be noted that all the resonator structures discussed above deliver extremely limited output power to the extent that absolute output power values are typically not reported. This low output also makes the characterisation of laser performance extremely challenging, leading to a debate in the community as to whether lasing has been actually achieved[18]. Although TMD resonant cavity emitters exhibit narrow linewidth, directionality, and polarisation of emission, many studies have faced challenges in convincingly demonstrating the expected two-fold linewidth reduction on threshold, spatial coherence, and output power, as previously mentioned. Additionally, the threshold behavior is typically subtle and can easily be mistaken for amplified spontaneous emission.

Here, we introduce a large area TMD laser based on a dual resonance metasurface for simultaneous absorption enhancement of the pump light and stimulation of the photoluminescence (PL) emission into a single lasing mode. By transferring a large area CVD-grown monolayer onto such a metasurface, we demonstrate a monolayer laser with 10s nW output power. We characterise the lasing behaviour and demonstrate remarkable temporal and spatial coherence that significantly exceeds the state-of-the-art in the field.

# Results

## CVD grown $WS_2$ monolayer

We choose $WS_2$ as the active material, not only because it has long-lived excitons[19, 20] and high gain[21] compared to other TMD gain materials, but also because high quality monolayer $WS_2$ can now be grown on a wafer scale[22, 23]. The homogeneity of PL from an as-grown $WS_2$ wafer is shown in Supplementary Information 1.

## Dual-resonance design - the rectangular lattice

To resonantly enhance both the absorption of the green pump laser (λ = 532 nm) and the PL emission of the $WS_2$, we designed a dielectric nanohole array, in $Si_3N_4$-on-glass substrate (Fig.1), with the period in x ($a_x$) suitably chosen to support a resonance for the pump wavelength and the period in y ($a_y$) to support a resonance at the emission wavelength (see Supplementary Information 2 for a detailed design and characterisation of the metasurface). We deliberately choose a mode with a relatively low Q-factor for the pump, in order to increase the angular tolerance required for the excitation with a focused laser beam; specifically, we chose a transverse-electric guided mode resonance (TE-GMR) mode, which results in a 2.6 fold enhancement in PL (simulation and experimental detail in Supplementary Information 2.3). On the other hand to achieve stimulated emission, we aim for the highest experimentally achievable Q-factor for a spatially extended resonant mode, to enable narrow-bandwidth lasing and low threshold operation. Two types of modes are available for this purpose, namely the transverse-magnetic guided-mode resonance mode ("TM-GMR") and the symmetry-protected bound-state-in-the-continuum TM mode ("TM-BIC") (Fig.2), both of which provide a high Q of ~ 3000 around the Γ-point (θ = 0°). The TM-BIC is a symmetry-protected mode with vanishing radiation loss at the Γ-point (more detail in Fig.2b and Supplementary Information 2). We will later show that this symmetry-protected BIC mode and the TM-GMR mode can both support lasing around the Γ-point.

The feedback mechanism in our design is provided by band-edge resonances associated with the aforementioned guided modes of the photonic crystals[24,25]. We note that although in an ideal structure (i.e., lossless and infinite size) the Q-factor at the Γ-point can be infinite in principle, in practical devices limited by scattering losses, the Q-factor of the TM-BIC mode becomes comparable to that of the TM-GMR mode (see Supplementary Information 5, Fig. S9).

The PMMA superstrate (Fig.1b) plays three important roles. Firstly, it acts as a carrier to transfer the $WS_2$ monolayer from its original growth substrate to the $Si_3N_4$ metasurface (see Methods). Second, it provides an advantageous mode overlap with the gain material because its refractive index ($n_{PMMA}$ = 1.49) is higher than air, thereby drawing the mode up towards the superstrate (Supplementary Information 4.2, Fig. S7). Third, the PMMA superstrate also results in a higher Q factor. Fourth, it protects the active material during the fabrication process, slows down its degradation and therefore prolongs the laser lifetime. In short, the PMMA layer is a critical component, as the absence of it results in the lack of any observed lasing activity.

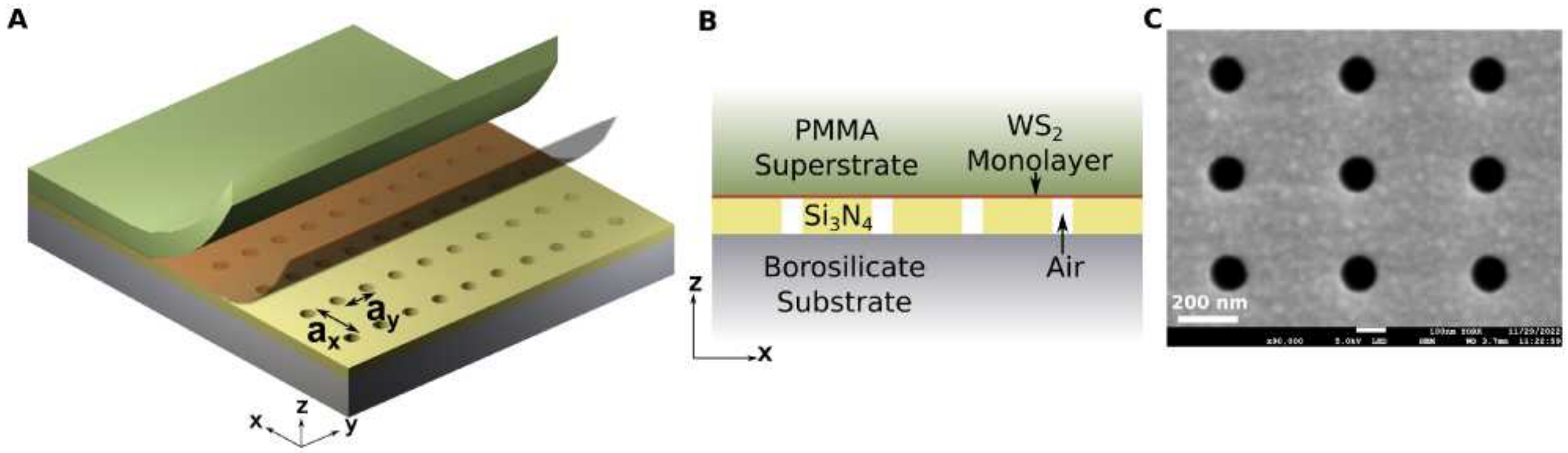

**Figure 1.** Dual-resonance nanohole array metasurface design and fabrication a) 3D-Rendering of the laser device, showing the rectangular lattice in $Si_3N_4$ with periods $a_x$ and $a_y > a_x$, the $WS_2$ and the PMMA layers. b) corresponding side view. c) SEM image of the fabricated bare $Si_3N_4$ metasurface (top view, x-y-plane). Periods $a_x$ = 410 nm and $a_y$ = 320 nm, with hole diameter $d$ = 140 nm.

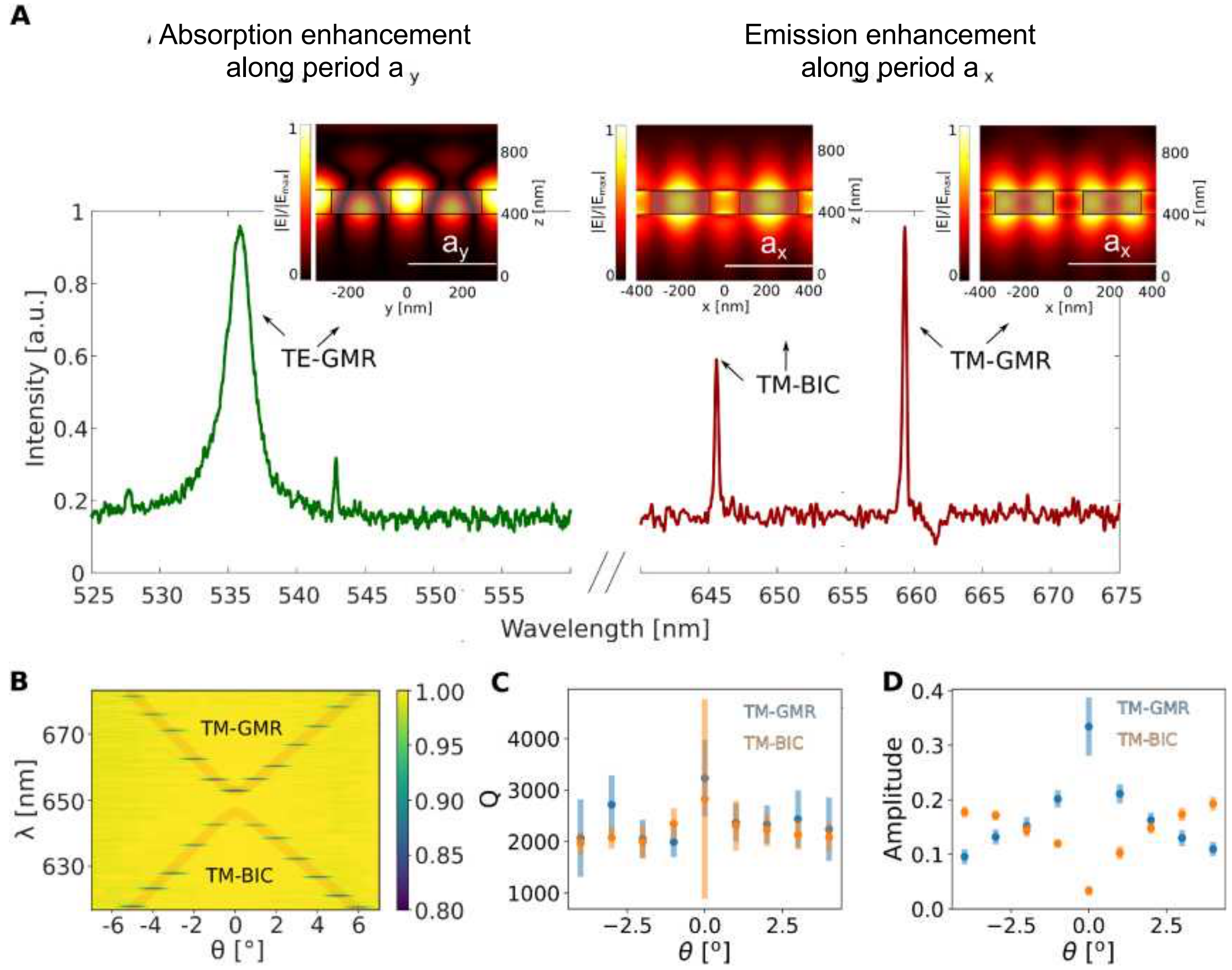

**Figure 2.** Experimental characterisation of the dual-resonance metasurface. a) Resonances supported by the rectangular lattice with the PMMA capping layer and without the $WS_2$ layer, in the wavelength ranges of excitation (green) and emission (red). The peak around 535 nm (blue curve) corresponds to a TE mode (dominant E-field component along x) while the peaks around 650 nm (red curve) correspond to the two TM modes, i.e. TM-GMR and TM-BIC with their field distributions plotted as insets. b) Band structure of the two available modes for emission enhancement around normal incidence with an angular resolution of $\Delta\theta = 1°$; measured in transmission. For the purpose of clearer visualisation, both bands are plotted in faint red lines corresponding to the TM-GMR (top) and TM-BIC (bottom) modes, showing the vanishing of the symmetry-protected BIC mode around $\theta = 0°$ and a photonic band gap at around 650 nm. c) Q-factor and d) corresponding amplitude for both TM modes extracted from the transmission measurements with an angular resolution of $\Delta\theta = 1°$. Note the error bar for the TM-BIC is very large at the Γ-point because of the weak signal.

## Laser characterisation: threshold, far-field radiation and spatial coherence

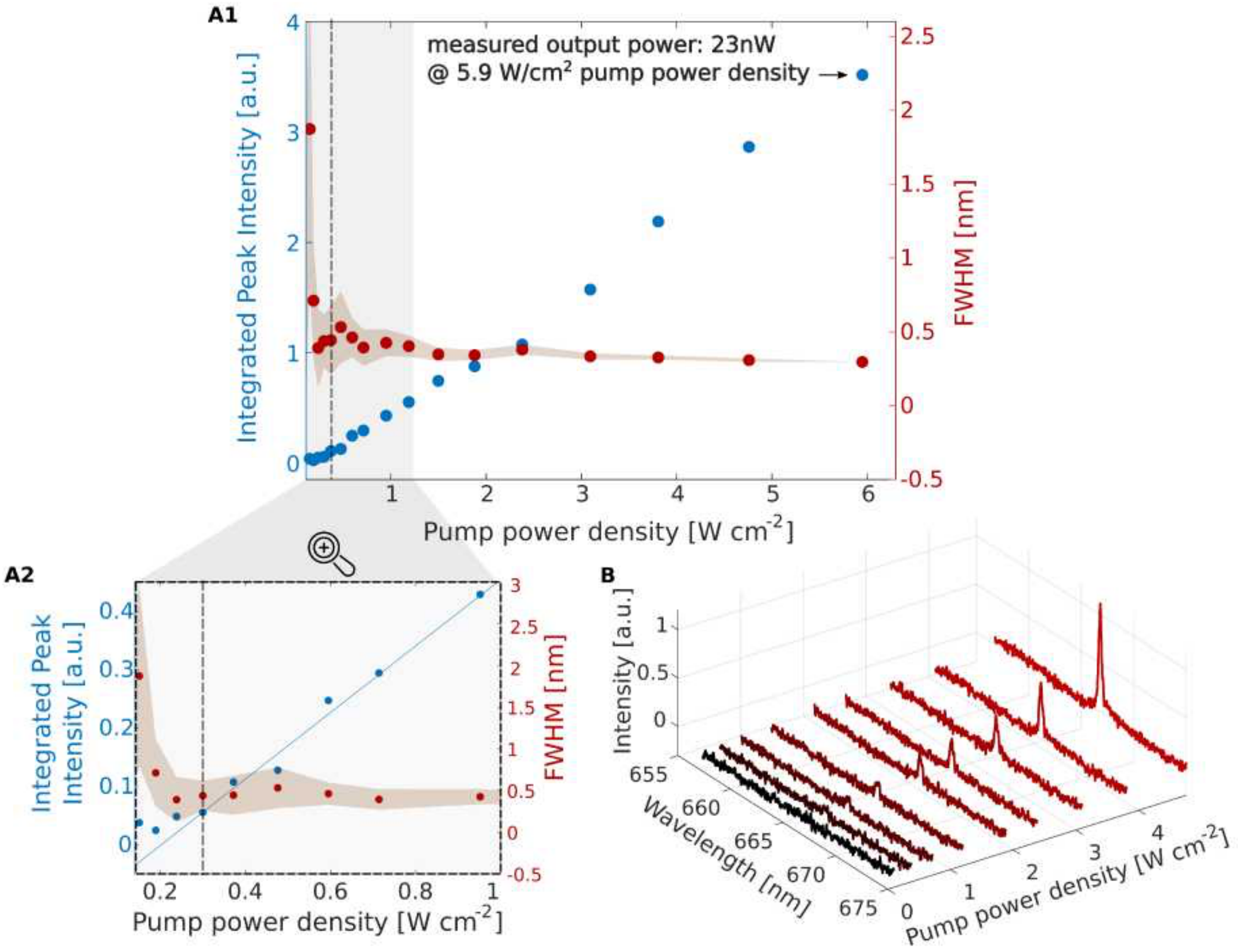

**Figure 3.** Laser threshold characterisation. a1) Example of the threshold behaviour (Light-light (L-L) curve with full width at half maximum (FWHM) as a function of pump intensity). a2) Zoom of a1) around the threshold (dashed line). The shaded region in the FWHM plot represents the 95% confidence interval of the fit. b) Emission spectra with different pump powers below and above threshold. The threshold of this device is at 0.3 W/cm$^2$.

We used lithographic tuning to establish the best combination of $a_x$ and $a_y$ values and constructed a matrix covering the range of $a_x$ = 400 to 425 nm and $a_y$ = 320 to 325 nm to match both the pump laser and the $WS_2$ emission wavelength. Out of the 15 devices characterised, we observed laser emission typically occurring at wavelengths above 645 nm, where the extinction coefficient of the $WS_2$ is sufficiently low to avoid the re-absorption of emitted photons (see Supplementary Information 9), in accordance with previous reports[9]. Low threshold densities and significant linewidth reductions were observed in all samples, see Figure 3 as an example. In terms of linewidth, we typically observed a reduction by a factor of 2-3, in line with established laser theory[26]. At low pump power, the signal-to-noise ratio is small, which introduces fitting uncertainty in that range. On the other hand, at high pump power levels, we measured an output power in excess of 20 nW. The measured output power is naturally lower than the total emitted power due to factors such as the collection lens with a numerical aperture of 0.1, which cannot capture all the light, and the presence of a 50/50 beam splitter in the collection path. As a result we estimate that the actual laser power is at least 100 nW (see Methods). Further discussion on estimated output power fulfilling the quantum threshold condition is included in the Supplementary Information 6.

The utilisation of both the rectangular lattice and the high refractive index superstrate discussed above is critical for achieving lasing at low pump power and for producing significant output power. Indeed, lasing was only observed with doubly resonant metasurfaces (i.e. no lasing was observed when the metasurface was not resonant at the pump wavelength). Another consequence of using a rectangular lattice instead of a square lattice is that it produces a line-shaped far-field emission pattern (Fig.4b and c). The observed shape is typical of lasers that exhibit feedback in a single direction. This is precisely the case with our rectangular lattice. The resonant emission enhancement is attained with a one-dimensional periodicity, resulting in a narrow laser beam along the x-direction of the lattice. In contrast, in the y-direction, the lasing mode can couple with a broad range of angles. Fig.4 shows lasing with the TM-GMR mode and the TM-BIC mode in the same device, along with their far-field emission patterns, respectively. Along the x direction, the divergence of the TM-BIC laser ($2\sigma = 8.8$ mrad) is approximately 2.5 times the divergence of the TM-GMR laser ($2\sigma = 3.5$ mrad). The larger divergence of the TM-BIC laser along the x-axis, compared to the TM-GMR laser, can be attributed to the inhibited radiation at the Γ-point. For both modes, the divergence angle along the y direction is significantly larger than the numerical aperture of our objective (NA = 0.1), which means that the measured output power value substantially underestimates the total emission power. The corresponding spectra below and above threshold are shown in Fig.4a.

To characterise the spatial coherence of our lasers, we use Young's slit method[27] by placing a double slit in the intermediate image plane and projecting the resulting interference pattern onto the camera (Fig. 5). Above the threshold, the interferograms show clear fringes with a visibility of at least 0.5 (see Fig.5 b and Methods) for a double slit spacing of 200 μm. Taking the magnification of the intermediate image plane into account, this spacing corresponds to a distance of 30 μm in the sample plane. We therefore conclude that the light emitted by the 2D material laser, for both the TM-GMR and the TM-BIC modes, is spatially coherent over distances of at least 30 μm, which is higher than reported for any comparable device[8, 28, 29].

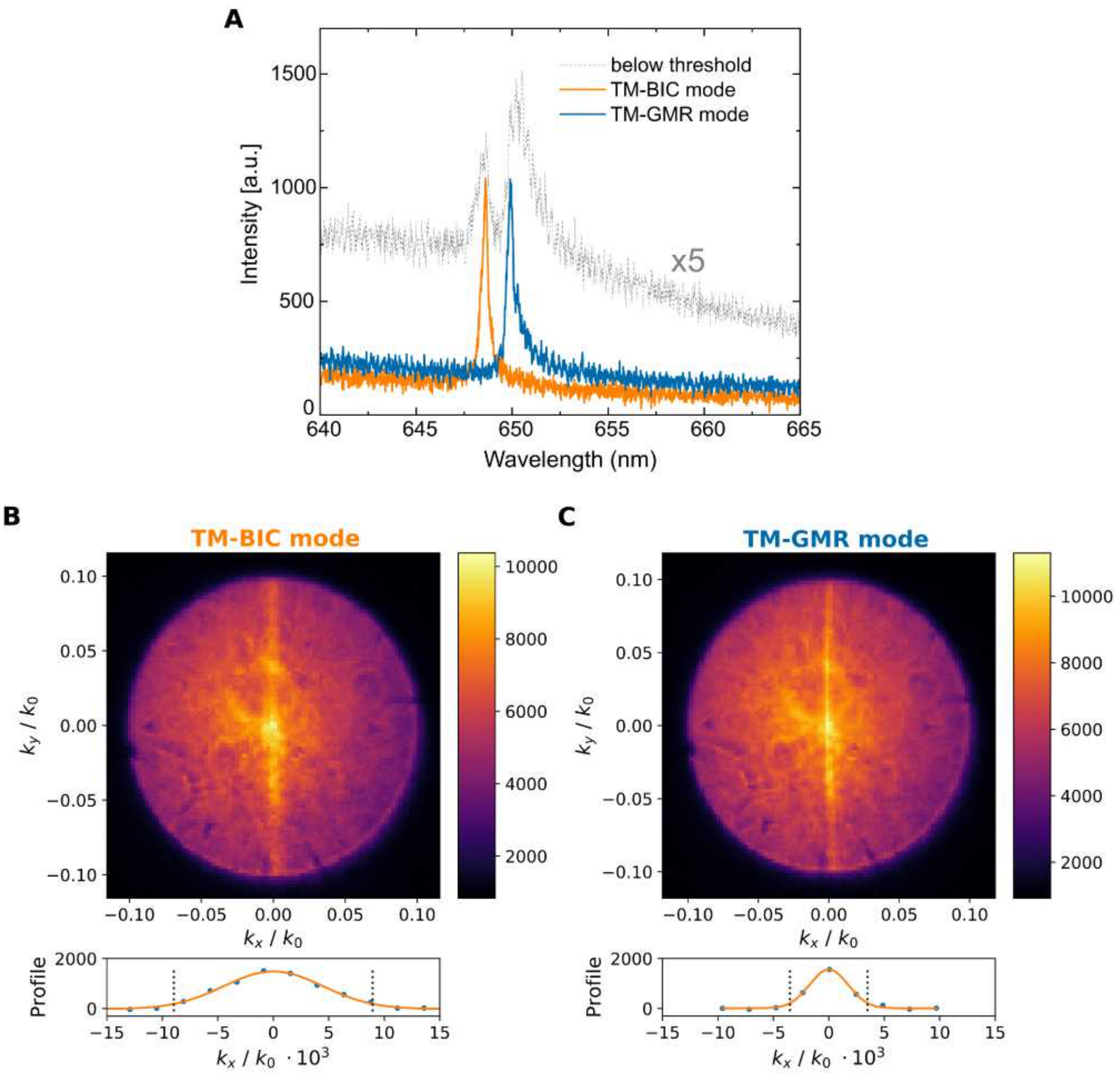

**Figure 4.** Lasing modes and their far-field patterns. a) Emission spectra below (x5) and above threshold, with lasing at the TM-BIC and the TM-GMR modes. b) and c) (top) Far-field emission patterns for the TM-BIC the TM-GMR modes, respectively. (bottom) An average profile along $k_x$ is obtained and baseline-corrected as described in Fig S11. Black dotted lines are plotted at $k_x = \pm 2\sigma$, with σ obtained from a Gaussian fit. More information can be found in Supplementary Information 8.

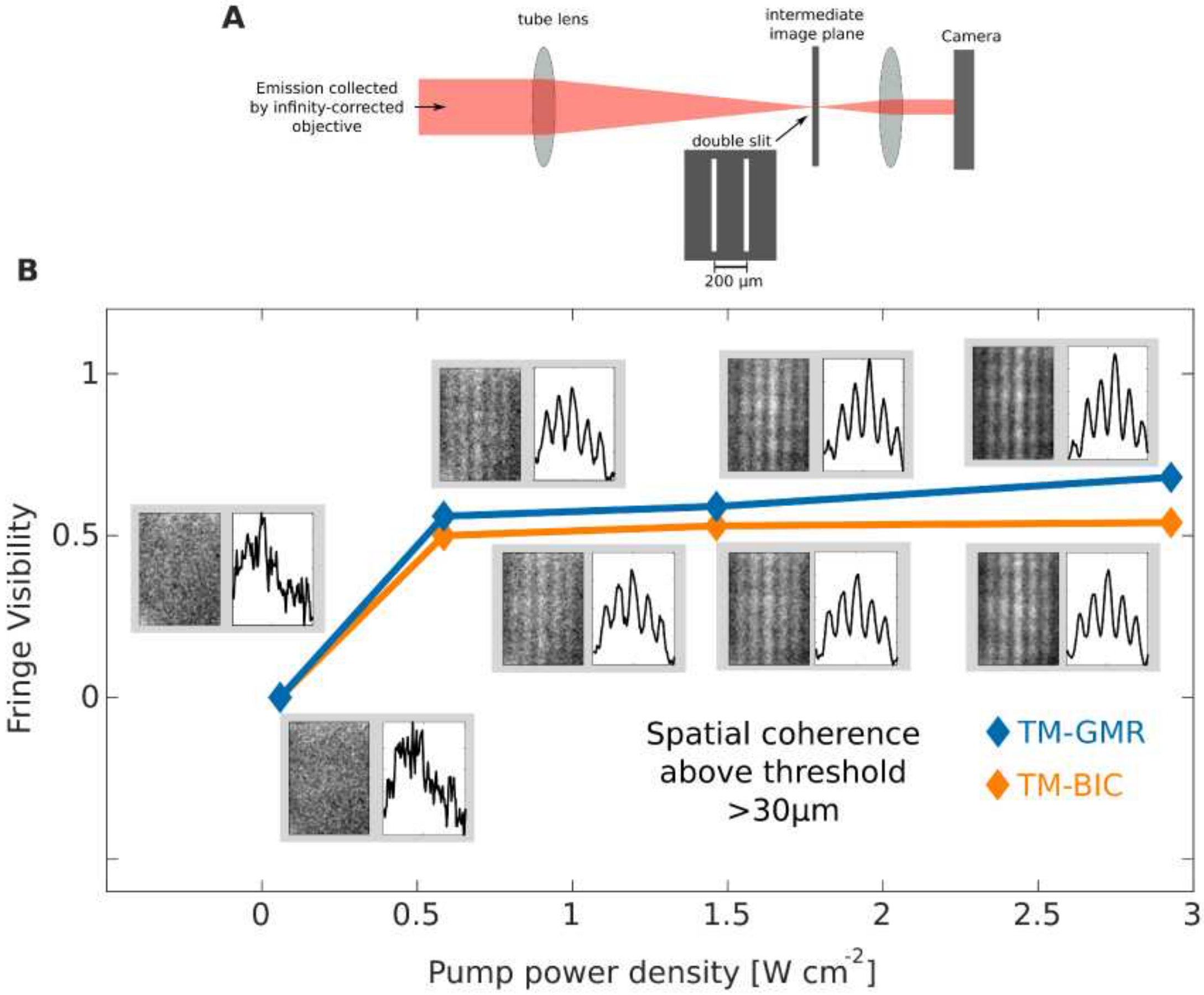

**Figure 5.** Characterisation of spatial coherence. a) Schematic of the double slit placed into an intermediate image plane (in addition to the micro-PL setup, see Supplementary Information 2.2) and b) Interferograms recorded below and above threshold with a double slit distance of 200 µm, corresponding to a distance of 30 µm on the sample. The interferograms were recorded with the same integration time at four different pump power values for both types of laser.

## Discussion

We have presented comprehensive evidence of lasing from a 2D material, enabled by a dielectric metasurface with a doubly resonant design. The rectangular array allows for the enhancement at both the pump and the emission wavelength of the laser, resulting in room temperature lasing with a remarkably low threshold of below 1 W/cm$^2$. Specifically, the lattice constant in one direction was chosen to resonate with the pump, while the lattice constant in the other direction supports high Q-factor (Q~3000) TM-polarised resonances to enhance laser emission. Notably, the two types of lasing modes we observed, i.e. the TM-GMR mode and the TM symmetry-protected BIC mode, exhibit similar threshold values and far-field emission patterns.

The resonant enhancement of both the pump and the PL emission, together with the large area of the laser, enabled by the wafer-scale MOCVD-grown $WS_2$ monolayer, allows us to obtain a very low threshold pump power density and an output power of tens of nW without damaging the active material (pumped at 20x the laser threshold). This relatively high output power was essential for a complete characterisation of the 2D material based lasing devices, especially concerning the far-field characterisation, the spatial coherence, and the quantum threshold condition. Notably, the threshold and output power we observe compare very favourably to other TMD devices (See Table 1, we present a selection of best performing devices reported in the literature. All characteristics were obtained at room temperature, expect for Ref.12, where lasing was achieved at at 130 Kelvin maximum). Our approach of absorption enhancement combined with emission enhancement in a rectangular lattice nanohole array, together with the large active area enabled by the extended guided-mode resonances and the MOCVD-grown material, is readily applicable to other light emitting TMDs and their heterostructures. Hence, the lasing operation can be readily extended to other wavelength ranges. Such dual resonances can also be utilized with other types of gain materials, including organic semiconductor thin films[30]. In conclusion, with our successful demonstration of practical emission powers and the potential for wafer-scale production, we anticipate that our research will open up new opportunities for the implementation of semiconductor light sources on heterogeneous substrates, such as flexible substrates/wearables. Additionally, our work holds promise for the advancement of biological and chemical sensing, as well as quantum optics, information, and other technology platforms.

**Table 1.** Laser performance comparison.

| Resonance category | Active material (" / " with capping) | Threshold ($W/cm^2$) | Pump Area ($\mu m^2$) | Extra characterisation$^+$ |
|---|---|---|---|---|
| Microdisk[10] | 4L $MoS_2$ | 630 | 0.785 | - |
| Defect-type cavity[12] | 1L $WSe_2$ | 1 | 2.69 | Reproducibility |
| VECSL[11] | 1L $WS_2$ | 0.442 | 1.13 | - |
| Nanobeam[31] | hBN/ 1L $MoTe_2$ / hBN | 6.6 | 4.67 | - |
| BIC[17] | 1L $WS_2$ / CYTOP | 144 | 0.95 | Far field emission |
| Microsphere (MP)[32] | 2L $WSe_2$ /MP | 0.72 | 1.56 | - |
| Our work | MOCVD 1L $WS_2$ / PMMA | 0.3 | 186265 | Far field emission and power; Spatial coherence; Tunability |

$^+$ Laser characterisation beyond the threshold behaviour

# Methods

## Design Simulations

For the design of the nanohole array[33], especially for identifying suitable periods to support the desired resonances, we simulated the resonances and their field components using the Rigorous Coupled Wave Analysis (RCWA) method[34]. For the refractive indices of the glass substrate, the metasurface material ($Si_3N_4$), and the PMMA polymer, we used $n_{glass}$ = 1.46[35], $n_{Si_3N_4}$ = 2.04[36] and $n_{PMMA}$ = 1.49[37], respectively. We assumed the values to be constant within the wavelength range of interest. We also carried out 3D FEM simulations (COMSOL Multiphysics) to verify the performance of the nanohole array metasurfaces. The results obtained from these simulations were consistent with the RCWA simulations. (See Supplementary Information 4).

## Metasurface Fabrication

We purchased glass wafers coated with silicon nitride (150 nm $Si_3N_4$ on 500 µm glass). These wafers were patterned by electron beam lithography (EBL) with a resist layer (AR-P 6200.13, Allresist GmbH) and a charge dissipation layer (AR-PC 5090, Allresist GmbH). We use an EBL system with an acceleration voltage of 50 kV (Voyager, Raith GmbH) to define the nanohole array pattern. We etch the pattern into the $Si_3N_4$ layer by means of reactive ion etching, using a mixed gas flow of $CHF_3$ and $O_2$ (29:1). In a final step, we remove the resist residue in 1165, a resist remover, leaving the metasurface ready for transferring the TMD gain material.

## Band Diagram of Cold Cavities

To characterise the cold cavity modes of the rectangular nanohole array before the $WS_2$ monolayer is applied, we spin-coat the sample with a 400 nm thick PMMA (Poly(methyl methacrylate)) layer. The band structure is measured in transmission on a rotation stage with a collimated visible light source and a high-resolution spectrometer (Acton Spectrapro 2750). The transmission spectra are acquired by tilting the sample across a range of angles with respect to the beam. The resulting spectra are normalised by the thin film response of the bare $Si_3N_4$ substrate.

## $WS_2$ Laser Fabrication

Monolayer $WS_2$ is grown on a 2-inch sapphire substrate by AIXTRON Ltd and AIXTRON SE. using a Close Coupled Showerhead metal-organic vapour deposition reactor. The uniformity of as-grown ML $WS_2$ across the 2-inch wafer is assessed using photoluminescence and Raman spectroscopy, as well as AFM and SEM. Centimetre-sized monolayers of $WS_2$ are lifted from the sapphire substrate with buffered oxide etching and then transferred from the sapphire substrate onto pre-patterned metasurfaces, using a wet-transfer process with a 400 nm thick PMMA layer as a carrier. The PMMA layer consequently acts as a superstrate and an encapsulation layer in the laser devices.

## Photoluminescence and Power Measurements

We use a micro-PL setup (Supplementary Information 2) to excite the devices with a CW laser (Novanta Photonics gem532) via a 4x objective (NA = 0.1) and collect the emission via the same objective. For the threshold characterisation, we record spectra with a range of pump powers. The excitation power on the sample is measured with a Thorlabs power meter (photodiode sensor model S121C). Various pump powers are obtained by inserting different optical neutral density (ND) filters in the excitation path. The excitation intensity is calculated with the reference power and the density of the ND filters. The emission power is measured after the 50/50 beamsplitter, with a silicon low power laser probe (Gentec-eo Pronto-Si, noise level 10 pW, the measured power being calibrated at 633 nm). The actual power from the laser device is estimated to be at least five times the measured power, taking into account the beam splitter, the objective NA and the the large beam divergence along the y-direction (the ’fan’ shape of the emitted beam).

## Spot Size of the Excitation Laser Beam

In order to determine the spot size of the excitation laser spot on the sample, we focus a real space image of the laser spot on the camera (CoolSNAP Myo) with a lens placed at its focal distance in front of the camera. The imaged spot is fitted with a 2D-Gaussian distribution, and the spot radius (radius = 77 µm) is taken as $r = 2a$, corresponding to the radius at which the intensity drops by $1/e^2$ of its peak value.

## Measurement of the Far Field Radiation Pattern

The far field radiation pattern was measured by imaging the back focal plane (BFP) of the objective onto the camera. To ensure that only the PL emission is captured, and the excitation light is blocked, we utilise a Thorlabs FEL0550 long pass filter. The axes of the resulting image of the BFP focused on the camera are proportional to the in-plane wave vector components $k_i = k_0 \sin(\theta_i)$, for $i \in x,y$. The pixel number along each axis is converted to $k_i$ by centering and multiplying with an appropriate calibration factor, which is found via the radius of the NA-circle.

## Acknowledgements

Y.W. acknowledges a Research Fellowship (TOAST, RF/201718/17131) awarded by the Royal Academy of Engineering. Z.H. and F.M. are partially supported by the Office of Naval Research with grant no. N00014-22-1-2486. T.F.K. acknowledges financial support from the Wellcome Trust (Funding ID 221349/Z/20/Z). G.S.A. and E.R.M. are supported by FAPESP (grants 2020/00619-4, 2020/159402, 2021/06121-0) and CNPq (307602/2021-4). We would like to acknowledge Prof. Yuri Kivshar at Australian National University for the insightful discussion on optical bound states in the continuum. We extend our acknowledgements to Haonan Tang at AIXTRON SE, for his help with PL analysis on as-grown wafers, and to Clifford McAleese and Ben Conran at AIXTRON Ltd, for their contribution to $WS_2$ MOCVD growth development.

## Author contributions statement

E.R. and Y.W. initiated the project. Y.W. fabricated the laser devices. I.B., M.D., D.C., G.S.A. and Z.H. performed modelling of the metasurfaces and the devices. I.B., M.D. and Y.W. performed the characterisation of the laser performance. S.P. and S.K. performed the $WS_2$ wafer growth and the initial characterisation of the wafer, M.H. oversaw the scientific development of 2D material wafer-scale growth at AIXTRON SE. All the authors analysed and discussed the data. I.B., M.D. and Y.W. wrote the manuscript with contributions from all co-authors.

## Data and code availability

The authors declare that all the data and code supporting the findings of this study are available within the article, or upon request from the corresponding author.

## Competing interests statement

The authors declare no competing interests.